\documentclass[aps,prd,floatfix,superscriptaddress,preprintnumbers,amsmath,amssymb,nofootinbib,showpacs,showkeys]{revtex4}
\usepackage{amssymb}
\usepackage{amsmath}
\usepackage{amsfonts}
\usepackage{graphicx}
\usepackage{epsfig}
\usepackage{xcolor}
\usepackage{xspace}
\usepackage{ulem}
\usepackage{lscape}
\usepackage{wasysym }
\usepackage[toc,page]{appendix}
\usepackage{mathtools}
\usepackage{multirow,rotating}
\usepackage{stackrel}
\usepackage{cancel}
\usepackage{pifont}
\newcommand{\eq}{\begin{eqnarray}}
\newcommand{\en}{\end{eqnarray}}

\def\gsim{\raise0.3ex\hbox{$\;>$\kern-0.75em\raise-1.1ex\hbox{$\sim\;$}}}
\def\lsim{\raise0.3ex\hbox{$\;<$\kern-0.75em\raise-1.1ex\hbox{$\sim\;$}}}

\begin{document}

\title{Deep inelastic $e-\tau$ and $\mu-\tau$ conversion \\
in the NA64 experiment at the CERN SPS}
\author{Sergei Gninenko} 
\affiliation{Institute for Nuclear Research, 117312 Moscow, Russia} 
\author{Sergey Kovalenko}
\affiliation{Departamento de F\'\i sica y Centro Cient\'\i fico
Tecnol\'ogico de Valpara\'\i so-CCTVal, Universidad T\'ecnica
Federico Santa Mar\'\i a, Casilla 110-V, Valpara\'\i so, Chile}
\author{Serguei Kuleshov}
\affiliation{Departamento de F\'\i sica y Centro Cient\'\i fico
Tecnol\'ogico de Valpara\'\i so-CCTVal, Universidad T\'ecnica
Federico Santa Mar\'\i a, Casilla 110-V, Valpara\'\i so, Chile}
\author{Valery E. Lyubovitskij}
\affiliation{Departamento de F\'\i sica y Centro Cient\'\i fico
Tecnol\'ogico de Valpara\'\i so-CCTVal, Universidad T\'ecnica
Federico Santa Mar\'\i a, Casilla 110-V, Valpara\'\i so, Chile}
\affiliation{Institut f\"ur Theoretische Physik, Universit\"at T\"ubingen, 
Kepler Center for Astro and \\ Particle Physics, 
Auf der Morgenstelle 14, D-72076, T\"ubingen, Germany}
\affiliation{Department of Physics, Tomsk State University, 
634050 Tomsk, Russia}
\affiliation{Tomsk State Pedagogical University, 634061 Tomsk, Russia}
\author{Alexey S. Zhevlakov}
\affiliation{Department of Physics, Tomsk State University,
634050 Tomsk, Russia}
\affiliation{Matrosov Institute for System Dynamics and Control Theory SB RAS, 
Lermontov str. 134, 664033, Irkutsk, Russia}



\begin{abstract} 

We study the Lepton Flavor Violating (LFV)  $e(\mu)-\tau$ conversion 
in Deep Inelastic Scattering (DIS) of electrons (muons) on 
fixed-target nuclei. Our model-independent analysis is based on 
the set of low-energy effective four-fermion LFV operators 
composed of leptons and quarks with the corresponding 
mass scales $\Lambda_{k}$ for each operator.
Using the estimated sensitivity of the search for this LFV process 
in events with large missing energy in the NA64 experiment 
at the CERN SPS, we derive lower limits for $\Lambda_{k}$ 
and compared them with the corresponding limits existing in the literature. 
We show that the DIS $e(\mu)-\tau$ conversion is able to provide a plenty 
of new limits as yet nonexisting in the literature. 
We also analyzed the energy spectrum of the final-state $\tau$ and 
discussed the viability of the observation of this process 
in the NA64 experiment and ones akin to it. The case of polarized beams 
and targets is also discussed.

\end{abstract}

\maketitle

\section{Introduction}
\label{sec:introduction}

The lepton flavor violation (LFV) is absent in the Standard Model (SM), 
if neutrinos are massless. Nowadays, nonzero neutrino masses and flavor 
mixing is a matter of experimental fact. 
LFV can be transmitted from the neutrino sector to the charged lepton one via 
the charged-current neutrino loop which, however, is heavily suppressed by 
the small neutrino mass square-differences. 
On the other hand, the possible high-scale physics beyond the SM (BSM) may 
contribute to the LFV in the charged lepton sector directly without 
the mediation of the neutrino sector. 
The high-scale BSM leaks into the low-energy theory via 
several universal effective nonrenormalizable LFV operators parametrizing 
in a generic way all the possible UV realization of BSM. 
In what follows, we specify these operators and estimate their possible 
contribution to the deep-inelastic $e-\tau$ and $\mu-\tau$ conversion 
\eq 
\label{eq:Proccess-1} 
e (\mu) + (A,Z)\rightarrow \tau + X
\en
of the initial electrons (muons) in the fixed-target with
the atomic and mass numbers $Z$ and $A$, respectively. 

The $e + p\rightarrow \tau + X$ process was searched for by 
the ZEUS Collaboration at HERA (DESY)~\cite{Chekanov:2002xz} 
in $e^+p$ collisions at a center-of-mass energy 
$\sqrt{s} \simeq$ 300 GeV. Theoretical study of 
the $e-\tau$, $\mu-e$, and  $\mu-\tau$ conversion has been done before 
in Refs.~\cite{Gninenko:2001id}-\cite{Takeuchi:2017btl}. 
The experimental study of $e-\tau$ and  $\mu-\tau$ conversion is planned 
by the NA64 experiment at the CERN SPS. 
The NA64 experiment is a fixed-target experiment combining 
the active beam dump and missing energy techniques to search for rare events.
The experiment will build and operate a fully hermetic detector placed on 
the H4 beam line at the CERN SPS with the primary goal to search for 
light dark photon ($A'$) coupled to photon, e.g. dark photons ($A'$),  
or sub-GeV dark gauge boson  $Z'$ coupled only to quarks or only to charged leptons. 
Other goals of the experiment are to search for the $L_\mu - L_\tau$  
gauge boson and the $K_L \to $ invisible decay,  
which is complementary to $K^+ \to \pi^+  \nu \bar\nu$,  
and invisible decays of light pseudoscalar mesons ($\pi^0$, $\eta$, $\etaœôó²$, $K_S$). 

The NA64 experiment is also capable for study
lepton conversion in the inclusive scattering of electrons or muons 
on nuclei $e^-(\mu^-) + (A,Z) \to \tau^- + X$.  This inclusive experimental mode 
is the only realistic for such kind of processes at the NA64 experimental setup. 
Note that elastic and quasi-elastic channels are included in deep inelastic
cross section we analyzed. For study of the $e-\tau$ conversion the NA64 experiment 
could employ the 100 GeV electron 
beam from the H4 beamline with a maximal intensity $\simeq (3-4) \cdot 10^6$ per SPS 
spill or 4.8 s produced by the primary 400 GeV/c proton beam with an intensity 
of a few $10^{12}$ protons on target. For study of the $\mu-\tau$ conversion 
it is planned to use the 150 GeV muon beam from the M2 beamline with the muon 
intensity $\simeq 2 \cdot 10^9$ per SPS spill and intesity 
of $10^{13}$ protons on target. 

\section{Theoretical Setup}

\label{sec:effoper}

We start with the low-energy effective Lagrangian 
relevant for two 
subprocesses of  $e-\tau$ (ETC) and $\mu-\tau$ (MTC) conversion:\\ 
(1) on quarks
\eq 
e^- + q_i \to \tau^- + q_f\,, \quad 
\mu^- + q_i \to \tau^- + q_f
\en
(2)  and on antiquarks 
\eq 
e^- + \bar q_f \to \tau^- + \bar q_i\,, \quad 
\mu^- + \bar q_f \to \tau^- + \bar q_i \,. 
\en 
Its most general form up to the dominant dim=6 operators is 
\eq 
\label{eq:eff-Lag-elltau}
\mathcal{L}_{\ell\tau} = \sum_{I,if,XY} \, 
\left(\Lambda^{\ell\tau}_{I_{if,XY}}\right)^{-2}  
\, \mathcal{O}^{\ell\tau}_{I_{if,XY}} \, + \, {\rm H.c.}
\,, \quad \ell = e, \mu\,, 
\en                          
where 
\eq 
\label{eq:eff-operators-S}
\mathbf{S}-\mbox{type:} &&\mathcal{O}^{\ell\tau}_{S_{if,XY}} =
(\bar\tau P_X l) 
(\bar q_f P_Y q_i)\,, \\
\label{eq:eff-operators-V}
\mathbf{V}-\mbox{type:} &&\mathcal{O}^{\ell\tau}_{V_{if,XY}}= 
(\bar\tau \gamma^\mu P_X l) 
(\bar q_f \gamma_\mu P_Y q_i)\,, \\
\label{eq:eff-operators-T}
\mathbf{T}-\mbox{type:} &&\mathcal{O}^{\ell\tau}_{T_{if,XX}} = 
(\bar\tau \sigma^{\mu\nu} P_X l) 
(\bar{q_{f}} \sigma_{\mu\nu} P_X q_{i})\,
\en 
are dim=6 operators with $\ell = e^-, \mu^-$. 
In Eq.~(\ref{eq:eff-Lag-elltau}),  
the summation over $I=S,V,T$, the quark flavors 
$i, f = u, d, s, c, b, t$ and chiralities $X,Y = L,R$ are implied. 
As usual, $P_{L,R} = (1\mp \gamma_{5})/2$ are the chirality projection 
operators. 
The mass scales $\Lambda^{\ell\tau}_{I_{if,XY}}$ set 
the strength of the low-energy effect of the corresponding operators. 
In total there are $360 = 6 \times 6 \times 10$ operators for the six quark 
flavors for each quark field and ten possible chirality combinations. 
These operators are subject to various already existing experimental 
constraints. Using these constraints we estimate the physics reach of  
the NA64 experiment~\cite{NA64} in the sense of the prospects  
of the observation of ETC and MTC~(\ref{eq:Proccess-1}) 
or improving the existing limits on the scales $\Lambda^{\ell\tau}_{I_{if,XY}}$.   
Prospects for the experimental searching for $\mu-\tau$ conversion  
in different context have been previously discussed 
in Refs.~\cite{Gninenko:2001id,Gninenko:2004wq,Black:2002wh} 
for the scalar operators.  

Note that we use nonuniversal scales $\Lambda^{\ell\tau}_{AB}$ 
to characterize the strength of the corresponding low-energy effective 
pointlike operators. From the view point of a high-scale underlying theory 
the operators~(\ref{eq:eff-operators-S})-(\ref{eq:eff-operators-T}) 
represent low-energy limits of the diagrams with two renormalizable vertices and 
a heavy intermediate particle of a typical mass $\mathcal{M}_{0}$. 
These diagrams are proportional to a product of two coupling constants, say,  
$g_{a}$ and $g_{b}$. We denote this product $\mathcal{C}^{\ell\tau}_{AB}\equiv g_{a}g_{b}$.  
After integrating out the heavy particles each operator goes accompanied with the factor 
\begin{eqnarray}
\label{eq:Coupl-Mass-Scale} 
\frac{\mathcal{C}^{\ell\tau}_{AB}}{M^{2}_{i}} \equiv 
\frac{1}{\left(\Lambda^{\ell\tau}_{AB}\right)^{2}}\,. 
\end{eqnarray} 
In what follows we derive lower limits on  $\Lambda^{\ell\tau}_{AB}$. 
With the above relation one may easily translate our limits to limits on 
$M_{i}$ for certain values of the effective couplings $\mathcal{C}^{\ell\tau}_{AB}$. 
The latter depend on a high-scale model.
For a weakly coupled high-scale model their ``natural'' 
values~\footnote{For more details we refer reader to Ref.~\cite{Black:2002wh}.} are
\begin{eqnarray}
\label{eq:nat}
\mathcal{C}^{\ell\tau}_{AB} = \mathcal{O}(1).
\end{eqnarray} 
Although the effective couplings can, in principle, be  significantly smaller we use in our analysis 
-- following the standard lore in the literature -- their ``natural'' values~(\ref{eq:nat}) 
in order to assure the validity of effective low-energy description in terms 
of pointlike effective operators. 

\section{Observables}
\label{sec:Observables}

First, we specify the kinematics. 
Let $P$, $p$, $p'$, $k$, and $k'$ be 
the momenta of initial nucleon, initial quark/final antiquark, 
final quark/initial antiquark, 
initial lepton, and final lepton, respectively. 
The set of invariant Mandelstam variables defining the kinematics of 
the quark/antiquark lepton scattering is given by 
\eq 
\hat{s} &=& (k + p)^2 = (k + x P)^2 \,, \nonumber\\
\hat{t} &=& (k - k')^2 \,, \\
\hat{u} &=& (k - p')^2 \,, \nonumber 
\en 
obeying the condition $\hat{s} + \hat{t} + \hat{u} = 0$ for zero masses 
of quarks and nucleon in comparison with large value of initial lepton energy. 
Here $x$ is the Bjorken variable 
(the fraction of the nucleon momentum carried by $q_{i}$ or $\bar{q}_{i}$):  
$x = Q^2/(q \cdot P)$.  
The inelasticity is 
$y = (q \cdot P)/(k \cdot P)$. The set $(\hat{s}, \hat{t}, \hat{u})$ 
is related to the total energy $s = (k + P)^2 \simeq 2 m_N E_\ell$, 
where $m_N$ is the nucleon mass and $E_\ell$ is the lepton beam energy 
\eq 
\label{eq:kinematcs}
\hat{s} &=& s x\,, \nonumber\\ 
\hat{t} &=& q^2 = - Q^2 = - s x y\,, \\ 
\hat{u} &=& - s x (1 - y)\,. \nonumber
\en 

\subsection{Integral cross section of the $\ell$-$\tau$ conversion} 
\label{sec:IntCross}

Now using effective four-fermion operators we calculate 
the integral cross sections for the ETC and MTC. 
The total cross section of the $l-\tau$ 
conversion on a nucleus~(\ref{eq:Proccess-1}) can be approximated 
by the sum over the corresponding cross section on its constituent nucleons
\eq 
\label{eq:xsect-1}
\sigma (\ell + (A,Z) \rightarrow \tau + X) &=& 
Z\  \sigma (\ell + p \rightarrow \tau + X) +
(A-Z) \  \sigma (\ell + n \rightarrow \tau + X).
\en
Here nucleon $N=p, n$ cross section is
\eq\label{eq:XSConversion-on-Nucleon}
\sigma (\ell + N \rightarrow \tau + X) &=& 
\sum\limits_{if} \, \int\limits_0^1  dx 
\int\limits_0^1  dy \ 
\biggl[
\frac{d^2\hat{\sigma}}{dx dy}(\ell + q_{i} \to 
\tau + q_{f}) \, q_{i}^{N}(x, Q^{2})\nonumber\\
&+&  
\frac{d^2\hat{\sigma}}{dx dy}(\ell + \bar q_{f} \to \tau + \bar q_{i})
\, \bar q_{f}^{N}(x, Q^{2}) 
\biggr] \, ,
\en 
where $q_{i}^{N}(x, Q^{2})$ and $\bar q_{i}^{N}(x, Q^{2})$ are 
quark and antiquark PDFs, respectively. 
We will consider two 
nuclear targets: Fe with $A=56$ and $Z=26$ and Pb with $A=207$ 
and $Z=82$. 
Quark/antiquark PDFs 
depend on the resolution scale set by the square momentum transferred  
to the nucleon
\eq\label{eq:deriv-01}
q^{2}= -Q^{2} = -(s - m_{N}^{2} - m_l^2) \, x y \simeq - s \, x \, y, 
\en
where $m_{N}$ is the nucleon mass, 
$x = Q^2/(q \cdot P)$ is Bjorken variable,
$y = (q \cdot P)/(k \cdot P)$ is inelasticity. 
Therefore, we should substitute $Q^{2}$ 
by $s x y $ in Eq.~(\ref{eq:XSConversion-on-Nucleon}). 
In the present paper we use quark PDFs from the 
CT10 next-to-next-to-leading order global analysis of QCD~\cite{Gao:2013xoa}. 
In fact, PDF fits using the standard 
CTEQ PDF evolution~\cite{Nadolsky:2008zw}  
but using the HOPPET $\alpha_s$ running solution. 

The elementary differential cross sections corresponding 
to the contact 4-fermion interactions 
in Eq.~(\ref{eq:eff-Lag-elltau}) are given by
\eq 
\label{eq:elementary-crosssections-q}   
\frac{d^2\hat{\sigma}}{dx dy}(\ell + q_{i} \to \tau + q_{f}) 
&=& \sum\limits_{I, XY} \, 
\frac{1}{\Big(\Lambda^{\ell\tau}_{I_{if,XY}}\Big)^4}
\frac{\hat{s} f_{I,XY}(y)}{64 \pi} \,, \\[3mm] 
\label{eq:elementary-crosssections-barq}   
\frac{d^2\hat{\sigma}}{dx dy}(\ell + \bar{q}_{f} \to \tau + \bar{q}_{i}) 
&=& \sum\limits_{I, XY} \, 
\frac{1}{\Big(\Lambda^{\ell\tau}_{I_{if,XY}}\Big)^4} 
\frac{\hat{s} g_{I,XY}(y)}{64 \pi} \,. 
\en
Here $f_{I,XY}(y)$ and $g_{I,XY}(y)$ are 
functions related to the matrix elements 
of the effective operators~(\ref{eq:eff-operators-S})-(\ref{eq:eff-operators-T}). 
They are given in Appendix~\ref{sec:Appendix}. 
 
Substituting~(\ref{eq:elementary-crosssections-q}), 
(\ref{eq:elementary-crosssections-barq}) into  
(\ref{eq:XSConversion-on-Nucleon}) and 
(\ref{eq:xsect-1}) we find
\eq 
\label{eq:XSConversion-on-Nucleus-1}
\sigma (\ell + (A,Z) \rightarrow \tau + X) &=& 
\sum\limits_{I, if, XY}
\, \frac{Q^A_{I_{if,XY}}}{\Lambda^{4}_{I_{if,XY}}}
\en
with
\eq 
\label{eq:QA-def}
Q^A_{I_{if,XY}} &=& 
\frac{s}{64 \pi}  \int\limits_0^1  dx 
\int\limits_0^1  dy \ \biggl[
x\, f_{I,XY}(y) \, q_{i}^{A}(x, s x y)
\, + 
\, x\, g_{I,XY}(y) \, \bar q_{f}^{A}(x, s x y) 
\biggr] \,,
\en
where
\eq 
\label{eq:qA_PDF}
&&u^A(x,Q^2) = Z u^p(x,Q^2) + (A-Z) d^p(x,Q^2)\,,\nonumber\\
&&d^A(x,Q^2) = Z d^p(x,Q^2) + (A-Z) u^p(x,Q^2)\,,\nonumber\\
&&u^A(x,Q^2) + d^A(x,Q^2) = A \ \Big(u^p(x,Q^2) + d^p(x,Q^2)\Big)\,,\nonumber\\
&&\bar u^A(x,Q^2) = A \bar u^p(x,Q^2)\,,\nonumber\\
&&\bar d^A(x,Q^2) = A \bar d^p(x,Q^2)\,,\nonumber\\
&&s^A(x,Q^2) = \bar s^A(x,Q^2) = A s^p(x,Q^2)\,,\nonumber\\
&&c^A(x,Q^2) = \bar c^A(x,Q^2) = A c^p(x,Q^2)\,,\nonumber\\
&&b^A(x,Q^2) = \bar b^A(x,Q^2) = A b^p(x,Q^2)  
\en 
are the quark and antiquark PDFs in a nucleus $A$. 
Numerical results for the double moments $Q^A_{I_{if,XY}}$ are shown 
in Tables~\ref{tab:t1}-\ref{tab:t4} for Fe and Pb nuclear targets and for
the electron and muon  beams.

The dominant contribution to the inclusive $\ell + A$ cross section 
is due to the bremsstrahlung of leptons on nuclei, given by the 
formula~\cite{Bethe:1934za,Tsai:1973py}  
\eq
\sigma_{BS} (\ell + (A,Z) \rightarrow \ell + X) = 4 \, \alpha \,
r_\ell^2 \, Z^2 \, \biggl[ \frac{7}{9} \, \log
  \biggl(\frac{183}{Z^{1/3}}\, \frac{m_\ell}{m_e}\biggr)\biggr] 
\en 
where $r_\ell = e^2/(4 \pi \epsilon_0 m_\ell c^2)$  
is the classical lepton radius: 
2.818 fm (for $e$) and 0.0136 fm (for $\mu$). 
For specific beam and target we have numerically 
\eq 
\sigma_{BS}(e   + Fe \rightarrow e + X) &=& 
0.129 \times 10^5 \ {\rm GeV}^{-2}\,, 
\nonumber\\
\sigma_{BS}(e   + Pb \rightarrow e + X) &=& 
1.165 \times 10^5 \ {\rm GeV}^{-2}\,, 
\nonumber\\
& &\\
\sigma_{BS}(\mu + Fe \rightarrow \mu + X) &=& 
0.692 \ {\rm GeV}^{-2}\,, 
\nonumber\\
\sigma_{BS}(\mu + Pb \rightarrow \mu + X) &=& 
6.607 \ {\rm GeV}^{-2}\,, 
\nonumber 
\en 
which will be used in the following section for the extraction of the limits 
on $\mu(e)-\tau$ LFV form the expected sensitivity of the NA64 experiment.

\subsection{Energy Spectrum}
\label{sec:Energy Spectrum}

An important characteristic helping to plan the $\ell-\tau$ conversion 
experiments is the energy spectrum of the final $\tau$-lepton 
defined as
\eq\label{eq:E-tau-Spectrum-1}
& &\mathcal{F}_{I}(E_{\ell},E_{\tau}) = 
\frac{1}{\sigma_{I}(l+(A,Z)\rightarrow \tau + X)}
\frac{d\sigma_{I}(l+(A,Z)\rightarrow \tau + X)}{d E_{\tau}},
\en
where $\sigma_{I}$ is the total cross section assuming 
the single operator $I = S,V,T$ dominance and the differential 
cross section is given by
\eq 
\frac{d\sigma_{I}}{dE_\tau} = 
\sum\limits_{if, XY} \, \frac{M_N}{32 \pi \Lambda^{4}_{I_{if,XY}}}
\int\limits_0^1  dx \, 
\biggl[ x \, q_{i}^{A}(x,\mu^2) \, f_{I,XY}(1-z) 
\,+\,   x \, \bar q_{f}^{A}(x,\mu^2) \, g_{I,XY}(1-z)
\biggr]
\en
with $\mu^2 = s x (1-z)$ and 
$z = E_\tau/E_\ell$ running from 0 to 1, which
corresponds to $0\leq E_\tau \leq E_\ell = s/(2 M_N)$.  
Here we use $d\hat{\sigma}/dE_\tau = - (2 M_N/s) \, (d\hat{\sigma}/dz)$. 
Note, the quantity $\mathcal{F}_{I}(E_{\ell},E_{\tau})$ is independent 
of the LFV scales  $\Lambda_{I}$. It is also independent of the target 
nucleus, since we sum over all the initial quark flavors $i$.
In Figs.~1-3 we plot the energy spectra $\mathcal{F}_{I}(E_{\ell},E_{\tau})$ 
for $I=S,V,T$ disregarding the quark 
contributions subdominant in comparison with $u+d$ quark and antiquark contribution. 
For simplicity, for each type of the LFV operator with specific spin structure 
$I=S,V,T,$ we suppose the same value of the coupling $\Lambda_I$ independent 
of the quark flavor. We use the following notations: 
$e_F$ and $\mu_F$ are the full contributions (including all species of quark and 
antiquarks) in case of the $E_e = 100$ GeV electron and 
                           $E_\mu = 150$ GeV muon beam, respectively; 
$e_{ud}$ and $\mu_{ud}$ are the respective $u+d$ contributions 
for the same values of energies of $e$ and $\mu$ beam. 

A useful ``integral'' quantity is the mean energy 
$\langle E_{\tau}\rangle$ of the final $\tau$ lepton defined as 
\eq\label{eq:METau-1}
\langle E_{\tau}\rangle_{I} &=& 
\int\limits_{0}^{E_{\ell}} d E_{\tau} \, E_{\tau} \,
\mathcal{F}_{I}(E_{\ell},E_{\tau}) \equiv   E_\ell \  
\frac{\sum\limits_{if,XY} \, \tilde Q^A_{I_{if,XY}}}{\sum\limits_{if,XY} \, 
Q^A_{I_{if,XY}}} \,,
\en 
where 
\eq
\tilde Q^A_{I_{if,XY}} = \frac{s}{64 \pi} \, 
\int\limits_0^1  dx x \, \int\limits_0^1 dz z \, 
\biggl[q_{i}^{A}(x,s x (1-z)) \, f_{I,XY}(1-z) 
\,+\, \bar q_{f}^{A}(x,s x (1-z))  \, g_{I,XY}(1-z)
\biggr]
\en 
separately for each of the operators $I=S,V,T$ 
in Eqs.~(\ref{eq:eff-operators-S})-(\ref{eq:eff-operators-T}). 
Here, as in case of Figs.~1-3  
for each type of the LFV operator with specific spin structure
$I=S,V,T$ we use the same value of the coupling $\Lambda_I$ independent
of the quark flavor. 
Note, that $E_{\tau}$ is independent of type of nucleus target 
because double moments of quark/antiquark 
$\tilde Q^A_{I_{if,XY}}$ and $Q^A_{I_{if,XY}}$ are both proportional to $A$. 

For the definition of $Q^{A}$ see Eq.~(\ref{eq:QA-def}). 
Our predictions for
$\tilde Q^A_{I_{if,XY}}$ and
$\langle E_{\tau}\rangle_{I}$ are displayed
in Tables~\ref{tab:tr1}-\ref{tab:tr4}
and in Tables~\ref{tab:Eaver1} and~\ref{tab:Eaver2}, respectively,
for $\ell = e, \mu$ beams and Fe, Pb targets.

For experiments searching for the LFV process~(\ref{eq:Proccess-1}),  
it is crucial that the missing energy in the decay of the final 
$\tau$-lepton be above some value. 
This is needed for the suppression of the typical backgrounds.
For the NA64 experiment, this cutoff is preliminarily estimated to be 
in the range 10 - 30 GeV (the detailed simulation results will be reported 
elsewhere). Thus, the $E_{\tau}$ should be large than the value.  
For example, for the cutoff of 10 GeV, the mean energy 
$\langle E_{\tau}\rangle_{I}$ of the final $\tau$-lepton is significantly 
larger than this value for the contribution of all the 
operators~(\ref{eq:eff-operators-S})-(\ref{eq:eff-operators-T})  
as one can seen from Tables~\ref{tab:Eaver1} and~\ref{tab:Eaver2}. 

\section{Limits on the LFV scales}
\label{sec:Limits on the LFV scales}

Here we derive limits on the mass scales $\Lambda$ of the LFV operators in 
Eqs.~(\ref{eq:eff-operators-S})-(\ref{eq:eff-operators-T}) and compare them 
with the corresponding limits existing in the literature.

\subsection{Expected limits from NA64 experiment}
\label{sec:Expected limits from NA64 experiment}

The quantity of the interest in the planning measurements of 
the electron (muon)-tau lepton conversion is the ratio:
\eq\label{eq:BR-1}
R_{\ell\tau} = \frac{\sigma(\ell + A \rightarrow \tau + X)}
{\sigma(\ell + A \rightarrow \ell  + X)}
\en
where $\sigma(\ell + A \rightarrow \ell  + X)\approx 
\sigma_{BS}(\ell + A \rightarrow \ell  + X)$. 

The physics reach of the NA64 experiment in this quantity is  
expected to be at the level of 
\eq 
\label{eq:NA64-reach}
&&R_{\ell\tau} \sim 10^{-13}-10^{-12}.
\en
Assuming single operator dominance,  the most optimistic value 
of~(\ref{eq:NA64-reach}) 
would result in the constraints on the LFV scales $\Lambda_{I_{if,XY}}$ 
shown in Tables~\ref{tab:tlow1}-\ref{tab:tlow4}.  
As seen from these tables, the limits are in the ranges 
\eq
\label{eq:OurLimSummary}
& & 
S-{\rm operators:} \  \Lambda^{e\tau} \ge 0.04 - 0.24 \ {\rm TeV}\,, \quad
\Lambda^{\mu\tau} \ge 0.56 - 3.05 \ {\rm TeV}\,, \nonumber\\
& &
V-{\rm operators:} \  \Lambda^{e\tau} \ge 0.05 - 0.44 \ {\rm TeV}\,, \quad
\Lambda^{\mu\tau} \ge 0.78 - 5.60 \ {\rm TeV}\,, \nonumber\\
& &
T-{\rm operators:} \  \Lambda^{e\tau} \ge 0.09 - 0.66 \ {\rm TeV}\,, \quad
\Lambda^{\mu\tau} \ge 1.45 - 10.06 \ {\rm TeV}\,. 
\en
The worst limits are set on the operators with $b$-quark. 
Such low scales may look incompatible with the concept of effective 
low-energy pointlike 4-fermion interactions 
(\ref{eq:eff-Lag-elltau})-(\ref{eq:eff-operators-T}) used in our analysis. 
In this respect the following comments are in order. 
First of all, the low values for the lower bounds of 
$\Lambda^{\ell\tau}_{AB}$ indicate deficiency of the method of their derivation 
saying nothing about their actual values. It is believed that the scales of 
the LFV operators are at a scale of $\gsim$1 TeV. Alas, for the weakly 
constrained scales, shown in (\ref{eq:OurLimSummary}), there are no experimental 
constraints as yet.  
Also, as we pointed out at the end of Sec.~\ref{sec:effoper}, we assume natural 
values (\ref{eq:nat}) of the effective couplings of a high-scale model underlying 
the effective low-energy Lagrangian~(\ref{eq:eff-Lag-elltau}). 
Then, for the masses of heavy LFV mediators we have $M_{0} \sim \Lambda^{\ell\tau}_{AB}$. 
Deviation from the point-like regime of the interactions 
(\ref{eq:eff-Lag-elltau}) would manifest itself in the propagator effect
\begin{eqnarray}
\label{eq:propagator-Effect} 
\frac{1}{M_{0}^{2}} \longrightarrow \frac{1}{\hat{t}+M_{0}^{2}} 
\end{eqnarray}
for the particle exchange in the $t$-channel and analogously in $u,s$ - channels. 
From Eqs.~(\ref{eq:kinematcs}) it follows for the average values  
$\langle \hat{t}\rangle,\langle \hat{s}\rangle,\langle \hat{u}\rangle 
\leq 94 (140)$ GeV$^{2}$ for the electron $E_{e} = 100$ GeV (muon $E_{\mu} = 150$ GeV) 
beam energy at NA64, which, for  $M_{0}^{2}\geq 1600$ GeV$^{2}$, is going to result 
in a mild propagator effect irrelevant for our rough estimations. 
An exception may happen for a Leptoquark (LQ) with the mass 
$M_{LQ}^{2}\leq 2 m_{N} E_{\ell}= 200 (300)$ GeV$^{2}$ in the $s$-channel for 
$E_{e}=100$ GeV ($E_{\mu}=150$ GeV). In this case, 
the pointlike picture breaks down and  
the on-shell LQ is produced in $\ell-q$-collision. However, the value $M_{LQ} \lsim 20$ GeV 
even for an LQ coupled only to $b-e$, for which there are neither direct nor indirect 
constraints, such a small value looks unlikely. 

Let us note that our limits shown in Tables~\ref{tab:tlow1}-\ref{tab:tlow4} 
have been derived from the best expected sensitivity~(\ref{eq:NA64-reach}) 
of the NA64 experiment and may look too optimistic. However, lower limits on $\Lambda$'s 
extracted from an experimental upper bound on $R_{l\tau}$  scale as  $(R_{l\tau})^{1/4} $ and, 
therefore, for the less optimistic case of $R_{\ell\tau} \sim10^{-12}$, or even worse,  
they will be comparable with the limits in our Tables 
and still valuable as they are obtained for the first time. 

\subsection{Limits from other experiments}
\label{sec:Limits from other experiments}

In the literature there exist limits on many of 
the operators~(\ref{eq:eff-operators-S})-(\ref{eq:eff-operators-T}). 
As a reference point, from the accelerator experiments let us mention 
the constraint from the ZEUS experiment~\cite{Chekanov:2002xz} 
at HERA (DESY), which is
\eq
\label{eq:HERA-lim}
\Lambda_{\rm ZEUS}^{e\tau} \ge 0.41 - 1.86 \ {\rm TeV}\,.
\en 
These limits apply to the scales for all the operators in 
Eq.~(\ref{eq:eff-operators-S})-(\ref{eq:eff-operators-T}), 
but only for the first generation quarks $u,d$.  
Note that Ref.~\cite{Chekanov:2002xz} used another basis of 
the four-fermion operators 
motivated by the leptoquark exchange, which we adjusted to our 
in Eqs.~(\ref{eq:eff-operators-S})-(\ref{eq:eff-operators-T}) and 
thus obtained (\ref{eq:HERA-lim}) from the limits on the leptoquark mass in
Ref.~\cite{Chekanov:2002xz}.

Limits on the operators~(\ref{eq:eff-operators-S})-(\ref{eq:eff-operators-T}) 
can be extracted from the decays, 
\eq\label{eq:decay-1}
&& \tau\rightarrow \ell + M^{0}\,,\nonumber\\
\label{eq:decay-2}
&&B\rightarrow \ell + \tau\,,\\
\label{eq:decay-3}
&&B\rightarrow \ell^{\pm} +\tau^{\mp} + M\,,
\nonumber 
\en
where $M$ is a generic meson allowed by energy-momentum conservation. 
Using the existing experimental bounds in Ref.~\cite{Black:2002wh},  
the authors extracted limits from some of these processes on the scales 
of the  scalar, pseudoscalar, vector and axial-vector 
$\mu-\tau$ LFV effective operators, which differ from our basis  
(\ref{eq:eff-operators-S})-(\ref{eq:eff-operators-T}). 
Thus, their limits translate to limits on linear combinations of our operators. 
Assuming no significant cancelations in these combinations we have
the results of Ref.~\cite{Black:2002wh} translated to the scales 
of our operators:
\eq 
\label{eq:Limits-decay-S}
\mathbf{S}-{\rm operators:}&& 
\Lambda_{S_{dd,XY}}\,, \Lambda_{S_{uu,XY}} \geq 5.8 \mbox{ TeV}\,,\  
\Lambda_{S_{ss,XY}} \geq 5 \mbox{ TeV}\,,\  
\Lambda_{S_{sd,XY}} \geq 1.9 \mbox{ TeV}\,,\  
\Lambda_{S_{bd,XY}} \geq 4.7 \mbox{ TeV}\,,\\[3mm] 
\label{eq:Limits-decay-V}
\mathbf{V}-{\rm operators:}&& 
\Lambda_{V_{dd,XY}}\,, \Lambda_{S_{uu,XY}} \geq 5.7 \mbox{ TeV}\,,\ 
\Lambda_{V_{ss,XY}} \geq 4.8 \mbox{ TeV}\,,\  
\Lambda_{V_{sd,XY}} \geq 1.8 \mbox{ TeV}\,\  
\Lambda_{V_{bd,XY}} \geq 4.1 \mbox{ TeV}\,. 
\en 
As seen from Tables~\ref{tab:tlow2} and~\ref{tab:tlow4},  
the DIS $\mu-\tau$ conversion on 
nuclei~(\ref{eq:Proccess-1}) covers a much wider set of 
the quark bilinears, providing limits on 
$\Lambda_{I_{if,XY}}$, where $i,f = u,d,s,c,b$ and $I=S,V,T$. 
Our limits shown in Tables~\ref{tab:tlow1}-\ref{tab:tlow4}
have been derived from the expected sensitivity~(\ref{eq:NA64-reach}) 
of the NA64 experiment. We conclude that the limits from 
the decays~(\ref{eq:decay-1})-(\ref{eq:decay-3}) on the set of 
$\Lambda$ in~(\ref{eq:Limits-decay-S})-(\ref{eq:Limits-decay-V}) 
are typically about factor stronger than ours, but there are missing limits 
on the operators with many other combinations of quark flavors, 
neither there are limits on the $\mathbf{T}$-operators, provided by 
the DIS $\mu-\tau$ conversion on nuclei~(\ref{eq:Proccess-1}). 

The following comment on the scalar operators may be in order.
It is known that the scalar LFV operators with heavy quarks can contribute LFV conversion 
via a triangle diagrams with two gluons legs immersed 
in a nucleus.  This contribution is significant for the case of  
coherent nuclear $\mu-e$-conversion 
(see, for instance, Refs.~\cite{Kosmas:2001sx,Cirigliano:2009bz}), 
being proportional to the nuclear mass.
For deep inelastic LFV conversion, studied in our paper, this contribution is 
the part of the full contribution of the heavy-quark nucleon sea and is  
characterized by the following stages
$g \rightarrow \bar{b} + b$,  $b + \ell_{1}\rightarrow \ell_{2} + b$,  
$\bar{b}+b \rightarrow g$.
In the inclusive setting of the experiment, which is the case of NA64, all the possible fates of 
the b quark are summed up after its interaction with the initial lepton, 
including channels with and without final $\bar{b}+b \rightarrow g$.
Obviously, the channel with annihilation in the final state is subdominant in this sum.  

Note that the $\mathbf{T}$-operators operators for some quark flavor 
combinations can be constrained by the decays~(\ref{eq:decay-3}). 
We have not found such limits in the literature and leave the study 
of this possibility for a future publication. 

\section{Polarized lepton beam and nuclear target}

Following Refs.~\cite{Bartelski:1979wc,Anselmino:1993tc}, we consider 
the lepton conversion for specific choice of spin configurations of 
incoming beam and nuclear target. 
For the studied energies of incoming leptons we may 
safely neglect quark and lepton masses. 
As shown in Refs.~\cite{Bartelski:1979wc,Anselmino:1993tc} for
$m_N \ll E_\ell$ the orientation of the nucleon spin $S$ is irrelevant, 
while there is a dependence on the helicity of the initial lepton $\lambda$. 
 
The differential cross section of the $\ell$-$\tau$ conversion,  
taking into account the initial lepton helicity $\lambda$, can be written as 
\eq 
\label{eq:Conversion-on-Nucleus-1_lambda}
\sigma_{\rm pol}(\ell + (A,Z) \rightarrow \tau + X) &=& 
\sum\limits_{I, if, XY}
\, \frac{Q^A_{I_{if,XY}}(\lambda)}{\Lambda^{4}_{I_{if,XY}}}
\en
with
\eq 
\label{eq:QAlambda}
Q^A_{I_{if,XY}}(\lambda) &=& 
\frac{s}{64 \pi}  \int\limits_0^1  dx 
\int\limits_0^1  dy \ \biggl[
x\, f_{I,XY}(y,\lambda) \, q_{i}^{A}(x, s x y)
\, + 
\, x\, g_{I,XY}(y,\lambda) \, \bar q_{f}^{A}(x, s x y) 
\biggr] \,,
\en
where the functions $f_{I,XY}(y,\lambda)$ and $g_{I,XY}(y,\lambda)$ 
are given in Appendix~\ref{sec:Appendix}. 

As seen from the Appendix, the choice of a particular helicity $\lambda$ of the
initial lepton allows one to eliminate the contributions of certain
operators to the $\ell-\tau$ flavor nondiagonal observables.
In particular, the contribution of the operators 
$\mathcal{O}^{\ell\tau}_{S_{if,RR}}$,  
$\mathcal{O}^{\ell\tau}_{S_{if,RL}}$,   
$\mathcal{O}^{\ell\tau}_{V_{if,RR}}$, 
$\mathcal{O}^{\ell\tau}_{V_{if,RL}}$, 
$\mathcal{O}^{\ell\tau}_{T_{if,RR}}$ is absent  
in quark-lepton subprocesses at $\lambda = +1$ 
and antiquark-lepton subprocesses at $\lambda = -1$. 
Correspondingly, the 
the contribution of the operators 
$\mathcal{O}^{\ell\tau}_{S_{if,LL}}$,  
$\mathcal{O}^{\ell\tau}_{S_{if,LR}}$,   
$\mathcal{O}^{\ell\tau}_{V_{if,LL}}$, 
$\mathcal{O}^{\ell\tau}_{V_{if,LR}}$, 
$\mathcal{O}^{\ell\tau}_{T_{if,LL}}$ 
is absent  
in quark-lepton subprocesses at $\lambda = -1$ 
and antiquark-lepton subprocesses at $\lambda = +1$. 
Therefore, the experiments with polarized lepton beams could be useful 
for separating the contributions of different LFV operators and in this way 
to help us identify the underlying LFV theory.
  
Note that the averaging of the polarized cross section  
in Eq.~(\ref{eq:Conversion-on-Nucleus-1_lambda})  
over $\lambda$ and $S$ reproduces our result  
for the unpolarized cross section (\ref{eq:XSConversion-on-Nucleus-1}),  
\eq 
\sigma(\ell + (A,Z) \rightarrow \tau + X) \equiv 
\frac{1}{4} \, \sum\limits_{\lambda, S} \, 
\sigma_{\rm pol}(\ell + (A,Z) \rightarrow \tau + X) 
\en 
and in agreement with Ref.~\cite{Anselmino:1993tc}. 

\section{Summary}

We presented phenomenological analysis of the $e-\tau$ and $\mu-\tau$ 
conversion in electron (muon) scattering on fixed-target nuclei (Pb and Fe). 
Our analysis is based on the effective four-fermion lepton-quark LFV operators 
linked to individual mass scales characterizing underlying renormalizable 
high-scale LFV physics. In the present paper, we have not considered its 
particular realization.

Using the expected sensitivity~(\ref{eq:NA64-reach}) 
of the NA64 experiment, we derived 
lower limits for the LFV mass scales of the effective operators 
and compared them with the corresponding limits existing in the literature, 
derived from the LFV decays. We have shown that the process 
of $e (\mu) + (A,Z)\rightarrow \tau + X$ is going to provide a plenty 
of new limits as yet nonexisting in the literature.

We predicted the spectrum of the final-state $\tau$-leptons helping one 
to assess the possibilities of discrimination of the signal from 
the typical backgrounds. We found that the mean $\tau$-lepton 
energy is significantly large than 
the required cutoff in the missing energy for all the effective 
operators~(\ref{eq:eff-operators-S})-(\ref{eq:eff-operators-T}).
In our opinion this result supports the viability of searching 
for the process~(\ref{eq:Proccess-1}) 
with the NA64 experiment, which will able to come up with 
new limits presented in Tables~\ref{tab:tlow1}-\ref{tab:tlow4}, 
many of which do not yet exist in the literature. 
Currently, the NA64 searching for invisible decay 
of dark photon does not see background
events with a large missing energy 
at the level $\sim 10^{-11}$ per 100 GeV electron 
on Pb target~\cite{na64-prd}. This result can be used to extract 
limits on $\Lambda$'s for the reaction of the $e-\tau$ conversion 
which will be reported elsewhere.  

We also examined the possible advantages of polarized beams and targets 
and showed that they can help distinguish the contributions of different 
LFV operators.  
We hope our results will be useful for planning experiments searching for 
Lepton Flavor Violation in accelerator experiments.

\begin{acknowledgments}

This work was funded 
by the German Bundesministerium f\"ur Bildung und Forschung (BMBF)
under Project 05P2015 - ALICE at High Rate (BMBF-FSP 202):
``Jet- and fragmentation processes at ALICE and the parton structure        
of nuclei and structure of heavy hadrons'',
by Fondecyt (Chile) Grant 
No.~1150792, and by CONICYT (Chile) Ring ACT1406, PIA/Basal FB0821. 
This research was supported by the Tomsk State University competitiveness
improvement program under Grant No. 8.1.07.2018. 

\end{acknowledgments}

\appendix
\section{Polarized and unpolarized matrix elements.}
\label{sec:Appendix}

Here we give the definitions and the explicit forms of 
the functions $f_{I}$ and $g_{I}$ in 
Eqs.~(\ref{eq:elementary-crosssections-q}), 
(\ref{eq:elementary-crosssections-barq}), and   
(\ref{eq:QAlambda}) involved in the calculation of 
the square matrix elements of 
the operators in Eqs.~(\ref{eq:eff-operators-S})-(\ref{eq:eff-operators-T}).

For the case of unpolarized $\ell-\tau$ conversion, we have 
\eq 
f_{S,LL}(y) &=& {\rm Tr}\Big[ P_L \not\! k  \ P_R \not\! k' \Big] 
             \  {\rm Tr}\Big[ P_L \not\! p' \ P_R \not\! p \Big] = y^2
\,, \nonumber\\ 
f_{S,RR}(y) &=& {\rm Tr}\Big[ P_R \not\! k  \ P_L \not\! k' \Big] 
             \  {\rm Tr}\Big[ P_R \not\! p' \ P_L \not\! p \Big] = y^2
\,, \nonumber\\ 
f_{S,LR}(y) &=& {\rm Tr}\Big[ P_L \not\! k  \ P_R \not\! k' \Big] 
             \  {\rm Tr}\Big[ P_R \not\! p' \ P_L \not\! p \Big] = y^2
\,, \nonumber\\ 
f_{S,RL}(y) &=& {\rm Tr}\Big[ P_R \not\! k  \ P_L \not\! k' \Big] 
             \  {\rm Tr}\Big[ P_L \not\! p' \ P_R \not\! p \Big] = y^2
\,, 
\en 
\eq 
f_{V,LL}(y) &=& {\rm Tr}\Big[ \gamma^\mu P_L \not\! k \ \gamma^\nu P_L \not\! k' \Big] 
             \  {\rm Tr}\Big[ \gamma_\mu P_L \not\! p \ \gamma_\nu P_L \not\! p' \Big] = 4
\,, \nonumber\\
f_{V,RR}(y) &=& {\rm Tr}\Big[ \gamma^\mu P_R \not\! k \ \gamma^\nu P_R \not\! k' \Big] 
             \  {\rm Tr}\Big[ \gamma_\mu P_R \not\! p \ \gamma_\nu P_R \not\! p' \Big] = 4
\,, \nonumber\\
f_{V,LR}(y) &=& {\rm Tr}\Big[ \gamma^\mu P_L \not\! k \ \gamma^\nu P_L \not\! k' \Big] 
             \  {\rm Tr}\Big[ \gamma_\mu P_R \not\! p \ \gamma_\nu P_R \not\! p' \Big] = 4 \, (1-y)^2
\,, \nonumber\\
f_{V,RL}(y) &=& {\rm Tr}\Big[ \gamma^\mu P_R \not\! k \ \gamma^\nu P_R \not\! k' \Big] 
             \  {\rm Tr}\Big[ \gamma_\mu P_L \not\! p \ \gamma_\nu P_L \not\! p' \Big] = 4 \, (1-y)^2
\,, 
\en 
\eq 
f_{T,LL}(y) &=& {\rm Tr}\Big[ \sigma^{\mu\nu} P_L \not\! k \ P_R \sigma^{\alpha\beta} \not\! k' \Big] \, 
                {\rm Tr}\Big[ \sigma_{\mu\nu} P_L \not\! p \ P_R \sigma_{\alpha\beta} \not\! p' \Big] = 16 \, (2-y)^2
\,, \nonumber\\
f_{T,RR}(y) &=& {\rm Tr}\Big[ \sigma^{\mu\nu} P_R \not\! k \ P_L \sigma^{\alpha\beta} \not\! k' \Big] \, 
                {\rm Tr}\Big[ \sigma_{\mu\nu} P_R \not\! p \ P_L \sigma_{\alpha\beta} \not\! p' \Big] = 16 \, (2-y)^2
\en 

and 
\eq 
g_{S,LL}(y) &=& {\rm Tr}\Big[ P_L \not\! k  \ P_R \not\! k'\Big] 
             \  {\rm Tr}\Big[ P_R \not\! p' \ P_L \not\! p \Big]= y^2
\,, \nonumber\\
g_{S,RR}(y) &=& {\rm Tr}\Big[ P_R \not\! k  \ P_L \not\! k' \Big] 
             \  {\rm Tr}\Big[ P_L \not\! p' \ P_R \not\! p \Big]= y^2
\,, \nonumber\\
g_{S,LR}(y) &=& {\rm Tr}\Big[ P_L \not\! k  \ P_R \not\! k' \Big] 
             \  {\rm Tr}\Big[ P_L \not\! p' \ P_R \not\! p \Big]= y^2
\,, \nonumber\\
g_{S,RL}(y) &=& {\rm Tr}\Big[ P_R \not\! k  \ P_L \not\! k' \Big] 
             \  {\rm Tr}\Big[ P_R \not\! p' \ P_L \not\! p \Big]= y^2
\,, 
\en 
\eq 
g_{V,LL}(y) &=& {\rm Tr}\Big[ \gamma^\mu P_L \not\! k  \ \gamma^\nu P_L \not\! k' \Big] \  
                {\rm Tr}\Big[ \gamma_\mu P_L \not\! p' \ \gamma_\nu P_L \not\! p \Big] = 4 \, (1-y)^2
\,, \nonumber\\ 
g_{V,RR}(y) &=& {\rm Tr}\Big[ \gamma^\mu P_R \not\! k  \ \gamma^\nu P_R \not\! k '\Big] \  
                {\rm Tr}\Big[ \gamma_\mu P_R \not\! p' \ \gamma_\nu P_R \not\! p \Big] = 4 \, (1-y)^2
\,, \nonumber\\ 
g_{V,LR}(y) &=& {\rm Tr}\Big[ \gamma^\mu P_L \not\! k  \ \gamma^\nu P_L \not\! k '\Big] \  
                {\rm Tr}\Big[ \gamma_\mu P_R \not\! p' \ \gamma_\nu P_R \not\! p \Big] = 4 
\,, \nonumber\\ 
g_{V,RL}(y) &=& {\rm Tr}\Big[ \gamma^\mu P_R \not\! k  \ \gamma^\nu P_R \not\! k '\Big] \ 
                {\rm Tr}\Big[ \gamma_\mu P_L \not\! p' \ \gamma_\nu P_L \not\! p \Big] = 4 
\,, 
\en 
\eq 
g_{T,LL}(y) &=& {\rm Tr}\Big[ \sigma^{\mu\nu} P_L \not\! k  \ \sigma^{\alpha\beta} P_R \not\! k' \Big] \  
                {\rm Tr}\Big[ \sigma_{\mu\nu} P_L \not\! p' \ \sigma_{\alpha\beta} P_R \not\! p \Big] 
= 16 \, (2-y)^2
\,, \nonumber\\ 
g_{T,RR}(y) &=& {\rm Tr}\Big[ \sigma^{\mu\nu} P_R \not\! k  \ \sigma^{\alpha\beta} P_L \not\! k' \Big] \ 
                {\rm Tr}\Big[ \sigma_{\mu\nu} P_R \not\! p' \ \sigma_{\alpha\beta} P_L \not\! p \Big] 
= 16 \, (2-y)^2
\,.
\en 
In the case of a polarized lepton beam and target, the corresponding functions 
are 
\eq 
f_{S,LL}(y,\lambda) &=& {\rm Tr}\Big[ P_L \not\! k  \, (1 + \gamma_5 \lambda) \ P_R \not\! k' \Big] 
                     \  {\rm Tr}\Big[ P_L \not\! p \ P_R \not\! p' \Big] = y^2 \, (1 + \lambda) 
\,, \nonumber\\ 
f_{S,RR}(y,\lambda) &=& {\rm Tr}\Big[ P_R \not\! k  \, (1 + \gamma_5 \lambda) \ P_L \not\! k' \Big] 
                     \  {\rm Tr}\Big[ P_R \not\! p \ P_L \not\! p' \Big] = y^2 \, (1 - \lambda) 
\,, \nonumber\\ 
f_{S,LR}(y,\lambda) &=& {\rm Tr}\Big[ P_L \not\! k  \, (1 + \gamma_5 \lambda) \ P_R \not\! k' \Big] 
                     \  {\rm Tr}\Big[ P_R \not\! p \ P_L \not\! p' \Big] = y^2 \, (1 +\lambda) 
\,, \nonumber\\ 
f_{S,RL}(y,\lambda) &=& {\rm Tr}\Big[ P_R \not\! k  \, (1 + \gamma_5 \lambda) \ P_L \not\! k' \Big] 
                     \  {\rm Tr}\Big[ P_L \not\! p \ P_R \not\! p' \Big] = y^2 \, (1 - \lambda) 
\,, 
\en 
\eq 
f_{V,LL}(y,\lambda) &=& {\rm Tr}\Big[ \gamma^\mu P_L \not\! k \, (1 + \gamma_5 \lambda) 
                         \gamma^\nu P_L \not\! k'\Big] 
                     \  {\rm Tr}\Big[ \gamma_\mu P_L \not\! p 
                     \ \gamma_\nu P_L \not\! p'\Big] = 4  \, (1 + \lambda)
\,, \nonumber\\
f_{V,RR}(y,\lambda) &=& {\rm Tr}\Big[ \gamma^\mu P_R \not\! k \, (1 + \gamma_5 \lambda) 
                         \gamma^\nu P_R \not\! k'\Big] 
                     \  {\rm Tr}\Big[ \gamma_\mu P_R \not\! p 
                     \ \gamma_\nu P_R \not\! p'\Big] = 4 \, (1 - \lambda)
\,, \nonumber\\
f_{V,LR}(y,\lambda) &=& {\rm Tr}\Big[ \gamma^\mu P_L \not\! k \, (1 + \gamma_5 \lambda) 
                         \gamma^\nu P_L \not\! k'\Big] 
                     \  {\rm Tr}\Big[ \gamma_\mu P_R \not\! p 
                     \ \gamma_\nu P_R \not\! p'\Big] = 4 \, (1-y)^2 \, (1 + \lambda)
\,, \nonumber\\
f_{V,RL}(y,\lambda) &=& {\rm Tr}\Big[ \gamma^\mu P_R \not\! k \, (1 + \gamma_5 \lambda) 
                         \gamma^\nu P_R \not\! k'\Big] 
                     \  {\rm Tr}\Big[ \gamma_\mu P_L \not\! p 
                     \ \gamma_\nu P_L \not\! p'\Big] = 4 \, (1-y)^2 \, (1 - \lambda)
\,, 
\en 
\eq 
f_{T,LL}(y,\lambda) &=& {\rm Tr}\Big[ \sigma^{\mu\nu} P_L \not\! k \, (1 + \gamma_5 \lambda) 
                          \sigma^{\alpha\beta} P_R \not\! k'\Big] 
                     \  {\rm Tr}\Big[ \sigma_{\mu\nu} P_L \not\! p 
                     \ \sigma_{\alpha\beta} P_R \not\! p'\Big] = 16 \, (2-y)^2 \, (1 + \lambda)
\,, \nonumber\\
f_{T,RR}(y,\lambda) &=& {\rm Tr}\Big[ \sigma^{\mu\nu} P_R \not\! k \, (1 + \gamma_5 \lambda) 
                          \sigma^{\alpha\beta} P_L \not\! k'\Big] 
                     \  {\rm Tr}\Big[ \sigma_{\mu\nu} P_R \not\! p 
                     \ \sigma_{\alpha\beta} P_L \not\! p'\Big] = 16 \, (2-y)^2 \, (1 - \lambda)\,, 
\en 
and 
\eq 
g_{S,LL}(y,\lambda) &=& {\rm Tr}\Big[ P_L \not\! k  \, (1 - \gamma_5 \lambda) \ P_R \not\! k' \Big] 
                     \  {\rm Tr}\Big[ P_L \not\! p' \ P_R \not\! p \Big] = y^2 \, (1 - \lambda) 
\,, \nonumber\\ 
g_{S,RR}(y,\lambda) &=& {\rm Tr}\Big[ P_R \not\! k  \, (1 - \gamma_5 \lambda) \ P_L \not\! k' \Big] 
                     \  {\rm Tr}\Big[ P_R \not\! p' \ P_L \not\! p \Big] = y^2 \, (1 + \lambda) 
\,, \nonumber\\ 
g_{S,LR}(y,\lambda) &=& {\rm Tr}\Big[ P_L \not\! k  \, (1 - \gamma_5 \lambda) \ P_R \not\! k' \Big] 
                     \  {\rm Tr}\Big[ P_R \not\! p' \ P_L \not\! p \Big] = y^2 \, (1 - \lambda) 
\,, \nonumber\\ 
g_{S,RL}(y,\lambda) &=& {\rm Tr}\Big[ P_R \not\! k  \, (1 - \gamma_5 \lambda) \ P_L \not\! k' \Big] 
                     \  {\rm Tr}\Big[ P_L \not\! p' \ P_R \not\! p \Big] = y^2 \, (1 + \lambda) 
\,, 
\en 
\eq 
g_{V,LL}(y,\lambda) &=& {\rm Tr}\Big[ \gamma^\mu P_L \not\! k \, (1 - \gamma_5 \lambda) 
                         \gamma^\nu P_L \not\! k'\Big] 
                     \  {\rm Tr}\Big[ \gamma_\mu P_L \not\! p' 
                     \ \gamma_\nu P_L \not\! p\Big] = 4 \, (1-y)^2 \, (1 - \lambda)
\,, \nonumber\\
g_{V,RR}(y,\lambda) &=& {\rm Tr}\Big[ \gamma^\mu P_R \not\! k \, (1 - \gamma_5 \lambda) 
                         \gamma^\nu P_R \not\! k'\Big] 
                     \  {\rm Tr}\Big[ \gamma_\mu P_R \not\! p' 
                     \ \gamma_\nu P_R \not\! p\Big] = 4 \, (1-y)^2 \, (1 + \lambda)
\,, \nonumber\\
g_{V,LR}(y,\lambda) &=& {\rm Tr}\Big[ \gamma^\mu P_L \not\! k \, (1 - \gamma_5 \lambda) 
                         \gamma^\nu P_L \not\! k'\Big] 
                     \  {\rm Tr}\Big[ \gamma_\mu P_R \not\! p' 
                     \ \gamma_\nu P_R \not\! p\Big] = 4 \, (1 - \lambda)
\,, \nonumber\\
g_{V,RL}(y,\lambda) &=& {\rm Tr}\Big[ \gamma^\mu P_R \not\! k \, (1 - \gamma_5 \lambda) 
                         \gamma^\nu P_R \not\! k'\Big] 
                     \  {\rm Tr}\Big[ \gamma_\mu P_L \not\! p' 
                     \ \gamma_\nu P_L \not\! p\Big] = 4 \, (1 + \lambda)
\,, 
\en 
\eq 
g_{T,LL}(y,\lambda) &=& {\rm Tr}\Big[ \sigma^{\mu\nu} P_L \not\! k \, (1 - \gamma_5 \lambda) 
                          \sigma^{\alpha\beta} P_R \not\! k'\Big] 
                     \  {\rm Tr}\Big[ \sigma_{\mu\nu} P_L \not\! p' 
                     \ \sigma_{\alpha\beta} P_R \not\! p\Big] = 16 \, (2-y)^2 \, (1 - \lambda)
\,, \nonumber\\
g_{T,RR}(y,\lambda) &=& {\rm Tr}\Big[ \sigma^{\mu\nu} P_R \not\! k \, (1 - \gamma_5 \lambda) 
                          \sigma^{\alpha\beta} P_L \not\! k'\Big] 
                     \  {\rm Tr}\Big[ \sigma_{\mu\nu} P_R \not\! p' 
                     \ \sigma_{\alpha\beta} P_L \not\! p\Big] = 16 \, (2-y)^2 \, (1 + \lambda)\,, 
\en 
where $\lambda$ is the helicity of the initial lepton.

\begin{table}[ht]
	\begin{center}
		\caption{Double moments of quark PDF 
                        $Q^A_{I_{if,XY}}$ (in GeV$^2$)
			with $f=u,d,s,c,b$ and $i$ specified in the Table. 
                        The case of
			a Fe target and an electron beam with $E_e = 100$ GeV.}
		\label{tab:t1}
		\def\arraystretch{1.}
		\begin{tabular}{|c|c||c|c|}
			\hline
			$(IiXY)$ & $Q^A_{I_{if,XY}}$ 
                        & $(IiXY)$ & $Q^A_{I_{if,XY}}$ \\
			\hline
			\hline
			\multicolumn{4}{|c|}{$S$ operators}\\
			\hline
			$(SuXY)$ &  3.82      & $(SdXY)$ & 4.07 \\
			$(SsXY)$ &  0.74      & $(ScXY)$ & 0.21 \\
			$(SbXY)$ &  0.006     &          &      \\
			\hline
			\multicolumn{4}{|c|}{$V$ operators}\\
			\hline
			$(VuLL/RR)$ &  43.83           & $(VuLR/RL)$ & 20.51 \\
			$(VdLL/RR)$ &  46.23           & $(VdLR/RL)$ & 22.46 \\
			$(VsLL/RR)$ &   5.85           & $(VsLR/RL)$ &  5.85 \\
			$(VcLL/RR)$ &   1.41           & $(VcLR/RL)$ &  1.41 \\
			$(VbLL/RR)$ &   0.02           & $(VbLR/RL)$ &  0.02 \\
			\hline
			\multicolumn{4}{|c|}{$T$ operators}\\
			\hline
			$(TuLL/RR)$ &  453.52       & $(TdLL/RR)$ & 484.37  \\
			$(TsLL/RR)$ &   81.84       & $(TcLL/RR)$ &  19.21  \\
			$(TbLL/RR)$ &    0.23       &             &         \\
			\hline
		\end{tabular}
	\end{center}
	
	\begin{center}
		\caption{The same as in Table~\ref{tab:t1}, but
			for a muon beam with $E_\mu = 150$ GeV.}
		\label{tab:t2}
		\def\arraystretch{1.}
		\begin{tabular}{|c|c||c|c|}
			\hline
			$(IiXY)$ & $Q^A_{I_{if,XY}}$ 
                        & $(IiXY)$ & $Q^A_{I_{if,XY}}$ \\
			\hline
			\hline
			\multicolumn{4}{|c|}{$S$ operators}\\
			\hline
			$(SuXY)$ &  5.64       & $(SdXY)$ & 6.01 \\
			$(SsXY)$ &  1.12       & $(ScXY)$ & 0.35 \\
			$(SbXY)$ &  0.02       &          &      \\
			\hline
			\multicolumn{4}{|c|}{$V$ operators}\\
			\hline
			$(VuLL/RR)$ &  64.30        & $(VuLR/RL)$ & 30.21 \\
			$(VdLL/RR)$ &  67.81        & $(VdLR/RL)$ & 33.07 \\
			$(VsLL/RR)$ &   8.84        & $(VsLR/RL)$ &  8.84 \\
			$(VcLL/RR)$ &   2.32        & $(VcLR/RL)$ &  2.32 \\
			$(VbLL/RR)$ &   0.07        & $(VbLR/RL)$ &  0.07 \\
			\hline
			\multicolumn{4}{|c|}{$T$ operators}\\
			\hline
			$(TuLL/RR)$ & 665.88        & $(TdLL/RR)$ & 710.69  \\
			$(TsLL/RR)$ & 123.48        & $(TcLL/RR)$ &  31.57  \\
			$(TbLL/RR)$ &   0.79        &             &         \\
			\hline
		\end{tabular}
	\end{center}
\end{table}

\begin{table}[ht]
	\begin{center}
		\caption{The same as in Table~\ref{tab:t1}, but for
			a Pb and an electron beam with $E_e = 100$ GeV.}
		\label{tab:t3}
		\def\arraystretch{1.}
		\begin{tabular}{|c|c||c|c|}
			\hline
			$(IiXY)$ & $Q^A_{I_{if,XY}}$ 
                        & $(IiXY)$ & $Q^A_{I_{if,XY}}$ \\
			\hline
			\hline
			\multicolumn{4}{|c|}{$S$ operators}\\
			\hline
			$(SuXY)$ & 13.58       & $(SdXY)$ & 15.61 \\
			$(SsXY)$ &  2.73       & $(ScXY)$ &  0.77 \\
			$(SbXY)$ &  0.02       &          &       \\
			\hline
			\multicolumn{4}{|c|}{$V$ operators}\\
			\hline
			$(VuLL/RR)$ & 155.11           & $(VuLR/RL)$ & 73.38 \\
			$(VdLL/RR)$ & 177.78           & $(VdLR/RL)$ & 85.47 \\
			$(VsLL/RR)$ &  21.63           & $(VsLR/RL)$ & 21.63 \\
			$(VcLL/RR)$ &   5.20           & $(VcLR/RL)$ &  5.20 \\
			$(VbLL/RR)$ &   0.08           & $(VbLR/RL)$ &  0.08 \\
			\hline
			\multicolumn{4}{|c|}{$T$ operators}\\
			\hline
			$(TuLL/RR)$ & 1610.52       & $(TdLL/RR)$ & 1856.30  \\
			$(TsLL/RR)$ &  302.50       & $(TcLL/RR)$ &   71.00  \\
			$(TbLL/RR)$ &    0.85       &             &          \\
			\hline
		\end{tabular}
	\end{center}
	
	\begin{center}
		\caption{The same as in Table~\ref{tab:t3}, but for
			a muon beam with $E_\mu = 150$ GeV.}
		\label{tab:t4}
		\def\arraystretch{1.}
		\begin{tabular}{|c|c||c|c|}
			\hline
			$(IiXY)$ & $Q^A_{I_{if,XY}}$ 
                        & $(IiXY)$ & $Q^A_{I_{if,XY}}$ \\
			\hline
			\hline
			\multicolumn{4}{|c|}{$S$ operators}\\
			\hline
			$(SuXY)$ & 20.03      & $(SdXY)$ & 23.02 \\
			$(SsXY)$ &  4.14      & $(ScXY)$ &  1.28 \\
			$(SbXY)$ &  0.07      &          &       \\
			\hline
			\multicolumn{4}{|c|}{$V$ operators}\\
			\hline
			$(VuLL/RR)$ & 227.61  & $(VuLR/RL)$ & 108.12 \\
			$(VdLL/RR)$ & 260.74  & $(VdLR/RL)$ & 125.80 \\
			$(VsLL/RR)$ &  32.67  & $(VsLR/RL)$ &  32.67 \\
			$(VcLL/RR)$ &   8.57  & $(VcLR/RL)$ &   8.57 \\
			$(VbLL/RR)$ &   0.25  & $(VbLR/RL)$ &   0.25 \\
			\hline
			\multicolumn{4}{|c|}{$T$ operators}\\
			\hline
			$(TuLL/RR)$ & 2365.21      & $(TdLL/RR)$ & 2723.18  \\
			$(TsLL/RR)$ &  456.45       & $(TcLL/RR)$ &   116.71  \\
			$(TbLL/RR)$ &    2.92       &             &          \\
			\hline
		\end{tabular}
	\end{center}
\end{table}

\begin{table}[ht]
	\begin{center}
		\caption{Double moments of quark PDF 
                        $\tilde Q^A_{I_{if,XY}}$ (in GeV$^2$)
			with $f=u,d,s,c,b$ and $i$ specified in the Table. 
                        The case of
			a Fe target and an electron beam with $E_e = 100$ GeV.
		}
		\label{tab:tr1}
		\def\arraystretch{1.}
		\begin{tabular}{|c|c||c|c|}
			\hline
			$(IiXY)$ & $\tilde Q^A_{I_{if,XY}}$ 
                        & $(IiXY)$ & $\tilde Q^A_{I_{if,XY}}$ \\
			\hline
			\hline
			\multicolumn{4}{|c|}{$S$ operators}\\
			\hline
			$(SuXY)$ &  0.97       & $(SdXY)$ & 1.03 \\
			$(SsXY)$ &  0.18       & $(ScXY)$ & 0.05 \\
			$(SbXY)$ &  0.001      &          &      \\
			\hline
			\multicolumn{4}{|c|}{$V$ operators}\\
			\hline
			$(VuLL/RR)$ &  23.08        & $(VuLR/RL)$ &  14.12 \\
			$(VdLL/RR)$ &  24.43        & $(VdLR/RL)$ &  15.31 \\
			$(VsLL/RR)$ &   3.28        & $(VsLR/RL)$ &  3.28  \\
			$(VcLL/RR)$ &   0.72        & $(VcLR/RL)$ &  0.72  \\
			$(VbLL/RR)$ &   0.005       & $(VbLR/RL)$ &  0.005 \\
			\hline
			\multicolumn{4}{|c|}{$T$ operators}\\
			\hline
			$(TuLL/RR)$ &  282.06       & $(TdLL/RR)$ & 301.39 \\
			$(TsLL/RR)$ &   49.55       & $(TcLL/RR)$ &  10.79 \\
			$(TbLL/RR)$ &    0.06       &             &        \\
			\hline
		\end{tabular}
	\end{center}
	
	\begin{center}
		\caption{The same as in Table~\ref{tab:tr1}, but for
			a muon beam with $E_\mu = 150$ GeV.}
		\label{tab:tr2}
		\def\arraystretch{1.}
		\begin{tabular}{|c|c||c|c|}
			\hline
			$(IiXY)$ & $\tilde Q^A_{I_{if,XY}}$ 
                        & $(IiXY)$ & $\tilde Q^A_{I_{if,XY}}$ \\
			\hline
			\hline
			\multicolumn{4}{|c|}{$S$ operators}\\
			\hline
			$(SuXY)$ &  1.42      & $(SdXY)$ & 1.52 \\
			$(SsXY)$ &  0.28      & $(ScXY)$ & 0.08 \\
			$(SbXY)$ &  0.003     &          &      \\
			\hline
			\multicolumn{4}{|c|}{$V$ operators}\\
			\hline
			$(VuLL/RR)$ & 33.82          & $(VuLR/RL)$ & 20.75 \\
			$(VdLL/RR)$ & 35.79          & $(VdLR/RL)$ & 22.48 \\
			$(VsLL/RR)$ &  4.94          & $(VsLR/RL)$ &  4.94 \\
			$(VcLL/RR)$ &  1.18          & $(VcLR/RL)$ &  1.18 \\
			$(VbLL/RR)$ &  0.02          & $(VbLR/RL)$ &  0.02 \\
			\hline
			\multicolumn{4}{|c|}{$T$ operators}\\
			\hline
			$(TuLL/RR)$ & 413.67      & $(TdLL/RR)$ & 441.89 \\
			$(TsLL/RR)$ &  74.58      & $(TcLL/RR)$ &  17.53  \\
			$(TbLL/RR)$ &   0.24      &             &         \\
			\hline
		\end{tabular}
	\end{center}
\end{table}

\begin{table}[ht]
	\begin{center}
		\caption{The same as in Table~\ref{tab:tr1}, but for
			for a Pb target and 
                        an electron beam with $E_e = 100$ GeV.}
		\label{tab:tr3}
		\def\arraystretch{1.}
		\begin{tabular}{|c|c||c|c|}
			\hline
			$(IiXY)$ & $\tilde Q^A_{I_{if,XY}}$ 
                        & $(IiXY)$ & $\tilde Q^A_{I_{if,XY}}$ \\
			\hline
			\hline
			\multicolumn{4}{|c|}{$S$ operators}\\
			\hline
			$(SuXY)$ &  3.44       & $(SdXY)$ & 3.96 \\
			$(SsXY)$ &  0.68       & $(ScXY)$ & 0.18 \\
			$(SbXY)$ &  0.004      &          &      \\
			\hline
			\multicolumn{4}{|c|}{$V$ operators}\\
			\hline
		$(VuLL/RR)$ &  81.73        & $(VuLR/RL)$ & 50.33 \\
		$(VdLL/RR)$ &  93.87        & $(VdLR/RL)$ & 58.45 \\
		$(VsLL/RR)$ &  12.13        & $(VsLR/RL)$ & 12.13 \\
		$(VcLL/RR)$ &   2.67        & $(VcLR/RL)$ &  2.67 \\
		$(VbLL/RR)$ &   0.02        & $(VbLR/RL)$ &  0.02 \\
		\hline
		\multicolumn{4}{|c|}{$T$ operators}\\
		\hline
		$(TuLL/RR)$ & 1001.40      & $(TdLL/RR)$ &  1155.30 \\
		$(TsLL/RR)$ &  183.18      & $(TcLL/RR)$ &   39.88 \\
		$(TbLL/RR)$ &   0.22       &             &         \\
		\hline
	\end{tabular}
\end{center}

\begin{center}
	\caption{The same as in Table~\ref{tab:tr3}, but for
		for a muon beam with $E_\mu = 150$ GeV.}
	\label{tab:tr4}
	\def\arraystretch{1.}
	\begin{tabular}{|c|c||c|c|}
		\hline
		$(IiXY)$ & $\tilde Q^A_{I_{if,XY}}$ 
                & $(IiXY)$ & $\tilde Q^A_{I_{if,XY}}$ \\
		\hline
		\hline
		\multicolumn{4}{|c|}{$S$ operators}\\
		\hline
		$(SuXY)$ & 5.06       & $(SdXY)$ &  5.82 \\
		$(SsXY)$ & 1.03       & $(ScXY)$ &  0.29 \\
		$(SbXY)$ & 0.02       &          &       \\
		\hline
		\multicolumn{4}{|c|}{$V$ operators}\\
		\hline
		$(VuLL/RR)$ &  119.80       & $(VuLR/RL)$ & 74.00 \\
		$(VdLL/RR)$ &  137.52       & $(VdLR/RL)$ & 85.79 \\
		$(VsLL/RR)$ &   18.26       & $(VsLR/RL)$ & 18.26 \\
		$(VcLL/RR)$ &   4.35        & $(VcLR/RL)$ &  4.35 \\
		$(VbLL/RR)$ &   0.07        & $(VbLR/RL)$ &  0.07 \\
		\hline
		\multicolumn{4}{|c|}{$T$ operators}\\
		\hline
		$(TuLL/RR)$ &  1469.25       & $(TdLL/RR)$ &1693.52  \\
		$(TsLL/RR)$ &  275.69      & $(TcLL/RR)$ &  64.82  \\
		$(TbLL/RR)$ &    0.89       &             &         \\
		\hline
	\end{tabular}
\end{center}
\end{table}

\begin{table}[ht]
\begin{center}
	\caption{$\langle E_\tau \rangle_I$ (in GeV)
		for different LFV operators in the case of
		an electron beam with $E_e = 100$ GeV.}
	\label{tab:Eaver1}
	\def\arraystretch{1.2}
	\begin{tabular}{|c|c|}
		\hline
		Operator       & $\langle E_\tau \rangle_I$ \\
		\hline
		$S$ operators  &  25  \\
		\hline
		$V$ operators  & 57  \\
		\hline
		$T$ operators  & 62    \\
		\hline
	\end{tabular}
\end{center}

\begin{center}
	\caption{The same as in Table~\ref{tab:Eaver1}, but for
		a muon beam with $E_\mu = 150$ GeV.}
	\label{tab:Eaver2}
	\def\arraystretch{1.2}
	\begin{tabular}{|c|c|}
		\hline
		Operator       & $\langle E_\tau \rangle_I$ \\
		\hline
		$S$ operators  &  38   \\
		\hline
		$V$ operators  &  86  \\
		\hline
		$T$ operators  &  93   \\
		\hline
	\end{tabular}
\end{center}
\end{table}

\begin{table}[ht]
	\begin{center}
		\caption{Lower limits for LFV mass scales
			$\Lambda_{I_{if,XY}}$ (in TeV) of the operators
			in Eq.~(\ref{eq:eff-Lag-elltau}) with $f=u,d,s,c,b$ and
			$i$ specified in the Table. The case of a Fe target and
			an electron beam with $ E_e = 100 $ GeV.}
		\label{tab:tlow1}
		\def\arraystretch{1.}
		\begin{tabular}{|c|c||c|c|}
			\hline
			$(IiXY)$ & $\Lambda_{I_{if,XY}}$ 
                        & $(IiXY)$ & $\Lambda_{I_{if,XY}}$ \\
			\hline
			\hline
			\multicolumn{4}{|c|}{$S$ operators}\\
			\hline
			$(SuXY)$ &  0.23      & $(SdXY)$ & 0.24 \\
			$(SsXY)$ &  0.15      & $(ScXY)$ & 0.11 \\
			$(SbXY)$ &  0.05      &          &      \\
			\hline
			\multicolumn{4}{|c|}{$V$ operators}\\
			\hline
			$(VuLL/RR)$ &  0.43           & $(VuLR/RL)$ & 0.35 \\
			$(VdLL/RR)$ &  0.44           & $(VdLR/RL)$ & 0.36 \\
			$(VsLL/RR)$ &  0.26           & $(VsLR/RL)$ & 0.26 \\
			$(VcLL/RR)$ &  0.18           & $(VcLR/RL)$ & 0.18 \\
			$(VbLL/RR)$ &  0.06           & $(VbLR/RL)$ & 0.06 \\
			\hline
			\multicolumn{4}{|c|}{$T$ operators}\\
			\hline
			$(TuLL/RR)$ &  0.77      & $(TdLL/RR)$ & 0.78  \\
			$(TsLL/RR)$ &  0.50      & $(TcLL/RR)$ & 0.34  \\
			$(TbLL/RR)$ &  0.12      &             &       \\
			\hline
		\end{tabular}
	\end{center}
	
	\begin{center}
		\caption{The same as in Table~\ref{tab:tlow1}, but for
			a muon beam with $E_\mu = 150$ GeV.}
		\label{tab:tlow2}
		\def\arraystretch{1.}
		\begin{tabular}{|c|c||c|c|}
			\hline
			$(IiXY)$ & $\Lambda_{I_{if,XY}}$ 
                        & $(IiXY)$ & $\Lambda_{I_{if,XY}}$ \\
			\hline
			\hline
			\multicolumn{4}{|c|}{$S$ operators}\\
			\hline
			$(SuXY)$ &  3.00      & $(SdXY)$ &  3.05 \\
			$(SsXY)$ &  2.01      & $(ScXY)$ &  1.50 \\
			$(SbXY)$ &  0.72      &          &      \\
			\hline
			\multicolumn{4}{|c|}{$V$ operators}\\
			\hline
			$(VuLL/RR)$ &  5.52           & $(VuLR/RL)$ & 4.57 \\
			$(VdLL/RR)$ &  5.60           & $(VdLR/RL)$ & 4.67 \\
			$(VsLL/RR)$ &  3.36           & $(VsLR/RL)$ & 3.36 \\
			$(VcLL/RR)$ &  2.41           & $(VcLR/RL)$ & 2.41 \\
			$(VbLL/RR)$ &  1.00           & $(VbLR/RL)$ & 1.00 \\
			\hline
			\multicolumn{4}{|c|}{$T$ operators}\\
			\hline
			$(TuLL/RR)$ &  9.90       & $(TdLL/RR)$ & 10.06  \\
			$(TsLL/RR)$ &  6.50       & $(TcLL/RR)$ & 4.62  \\
			$(TbLL/RR)$ &  1.84       &             &       \\
			\hline
		\end{tabular}
	\end{center}
\end{table}

\begin{table}[ht]
	\begin{center}
		\caption{The same as in Table~\ref{tab:tlow1}, but for
			a Pb target and an electron beam with $E_e = 100$ GeV.}
		\label{tab:tlow3}
		\def\arraystretch{1.}
		\begin{tabular}{|c|c||c|c|}
			\hline
			$(IiXY)$ & $\Lambda_{I_{if,XY}}$ 
                        & $(IiXY)$ & $\Lambda_{I_{if,XY}}$ \\
			\hline
			\hline
			\multicolumn{4}{|c|}{$S$ operators}\\
			\hline
			$(SuXY)$ &  0.18     & $(SdXY)$ & 0.19 \\
			$(SsXY)$ &  0.12     & $(ScXY)$ & 0.09 \\
			$(SbXY)$ &  0.04     &          &      \\
			\hline
			\multicolumn{4}{|c|}{$V$ operators}\\
			\hline
			$(VuLL/RR)$ &  0.34          & $(VuLR/RL)$ & 0.28 \\
			$(VdLL/RR)$ &  0.35          & $(VdLR/RL)$ & 0.29 \\
			$(VsLL/RR)$ &  0.21          & $(VsLR/RL)$ & 0.21 \\
			$(VcLL/RR)$ &  0.15          & $(VcLR/RL)$ & 0.15 \\
			$(VbLL/RR)$ &  0.05          & $(VbLR/RL)$ & 0.05 \\
			\hline
			\multicolumn{4}{|c|}{$T$ operators}\\
			\hline
			$(TuLL/RR)$ &  0.61       & $(TdLL/RR)$ & 0.63  \\
			$(TsLL/RR)$ &  0.40       & $(TcLL/RR)$ & 0.28  \\
			$(TbLL/RR)$ &  0.09       &             &       \\
			\hline
		\end{tabular}
	\end{center}
	
	\begin{center}
		\caption{The same as in Table~\ref{tab:tlow3}, but for
			a muon beam with $E_\mu = 150$ GeV.}
		\label{tab:tlow4}
		\def\arraystretch{1.}
		\begin{tabular}{|c|c||c|c|}
			\hline
			$(IiXY)$ & $\Lambda_{I_{if,XY}}$ 
                        & $(IiXY)$ & $\Lambda_{I_{if,XY}}$ \\
			\hline
			\hline
			\multicolumn{4}{|c|}{$S$ operators}\\
			\hline
			$(SuXY)$ &  2.35      & $(SdXY)$ & 2.45 \\
			$(SsXY)$ &  1.58      & $(ScXY)$ & 1.17 \\
			$(SbXY)$ &  0.56      &          &      \\
			\hline
			\multicolumn{4}{|c|}{$V$ operators}\\
			\hline
			$(VuLL/RR)$ &  4.31         & $(VuLR/RL)$ & 3.57  \\
			$(VdLL/RR)$ &  4.46          & $(VdLR/RL)$ & 3.71 \\
			$(VsLL/RR)$ &  2.65          & $(VsLR/RL)$ & 2.65 \\
			$(VcLL/RR)$ &  1.90          & $(VcLR/RL)$ & 1.90 \\
			$(VbLL/RR)$ &  0.78          & $(VbLR/RL)$ & 0.78 \\
			\hline
			\multicolumn{4}{|c|}{$T$ operators}\\
			\hline
			$(TuLL/RR)$ &  7.74      & $(TdLL/RR)$ & 8.01  \\
			$(TsLL/RR)$ &  5.13      & $(TcLL/RR)$ & 3.65  \\
			$(TbLL/RR)$ &  1.45      &             &       \\
			\hline
		\end{tabular}
	\end{center}
\end{table}

\newpage 

\begin{figure}[htb]
\begin{center}

	\epsfig{figure=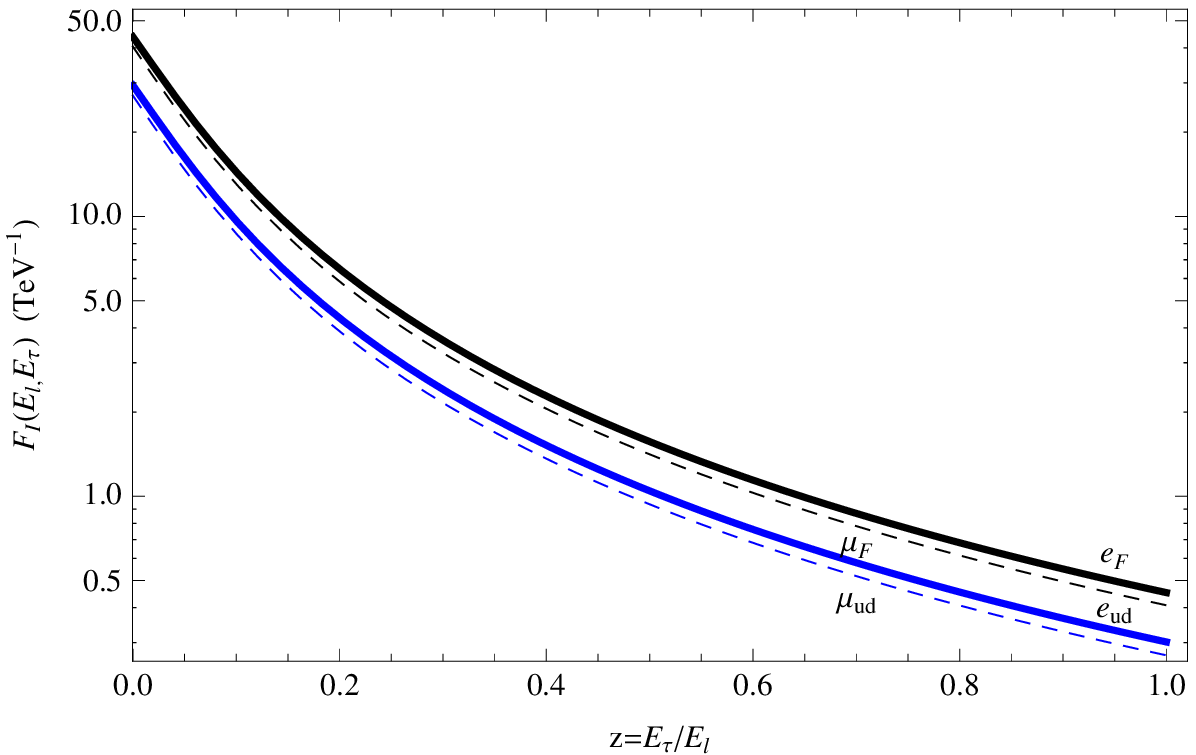,scale=0.8}
	\caption{Spectrum of $\tau$ lepton for $S$ operators.} 

\vspace*{.25cm}

	\epsfig{figure=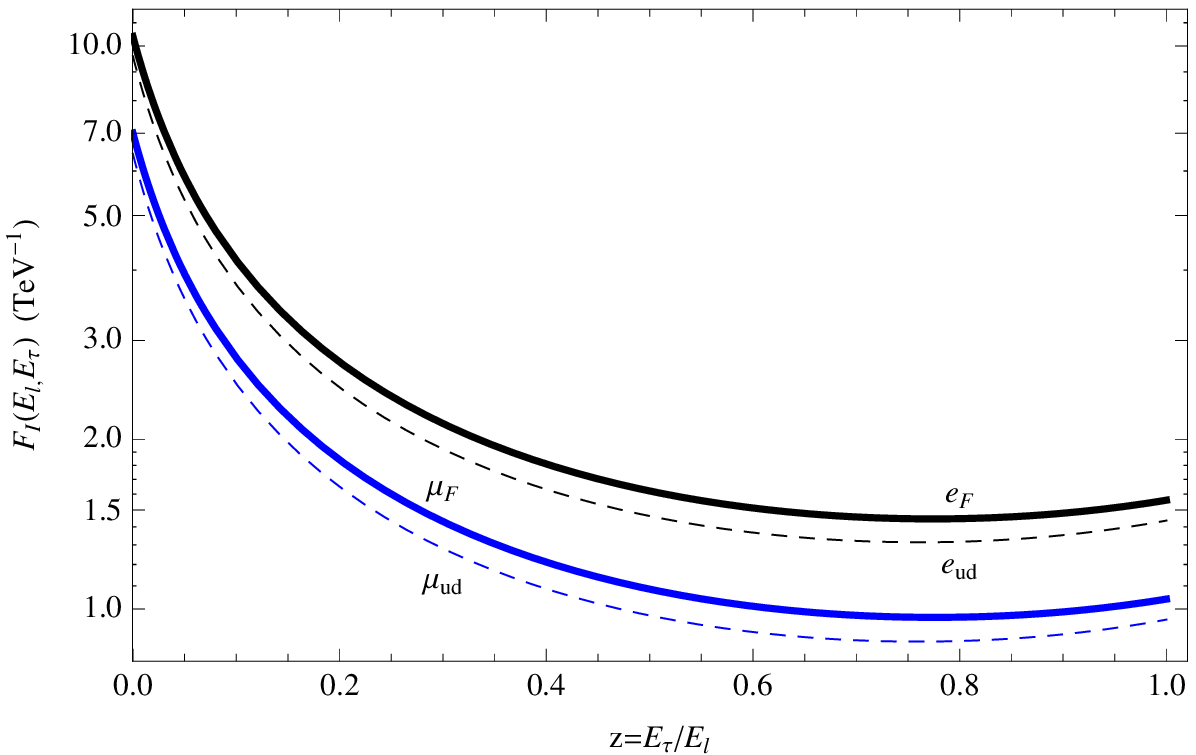,scale=0.8}
	\caption{Spectrum of $\tau$ lepton for $V$ operators.}

\vspace*{.25cm}

	\epsfig{figure=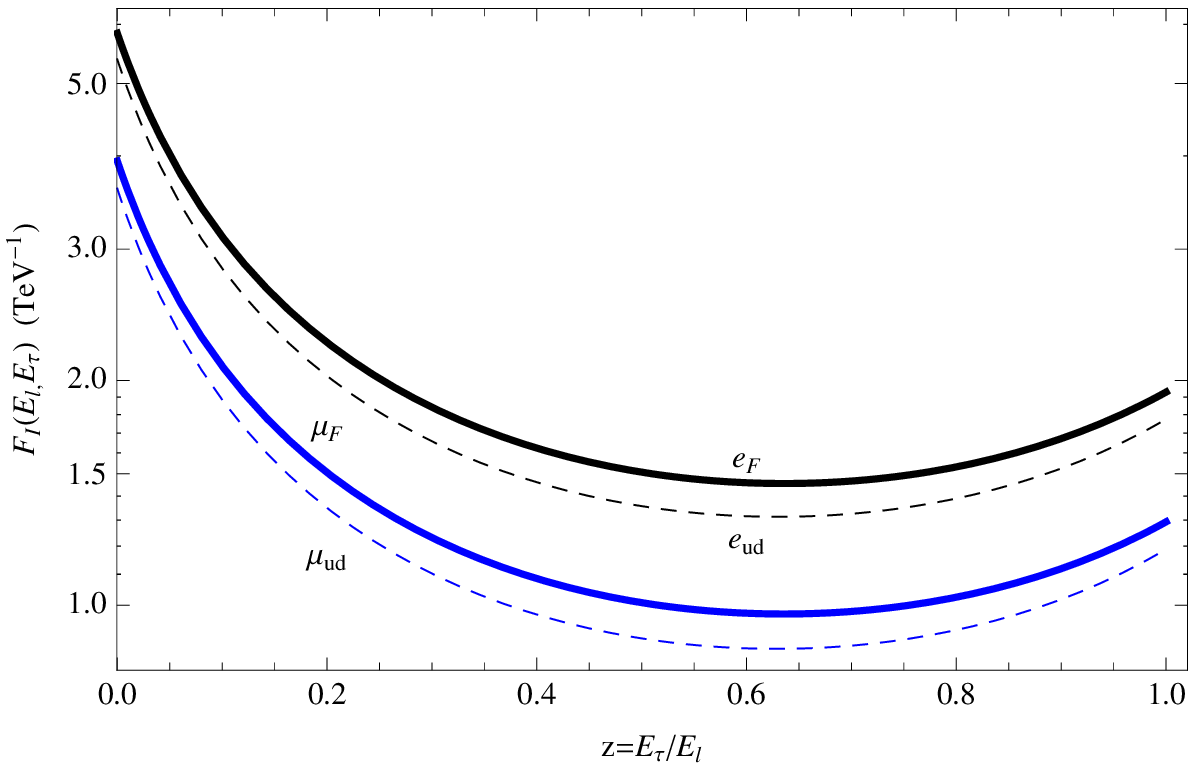,scale=0.8}
	\caption{Spectrum of $\tau$ lepton for $T$ operators.}

\end{center}
\end{figure}

\end{document}